# On-chip Optical Phase Monitoring in Multi-Transverse-Mode Integrated Silicon-based Optical Processors

Kaveh (Hassan) Rahbardar Mojaver, *Member, IEEE*, Odile Liboiron-Ladouceur, *Senior Member, IEEE*

*Abstract*—We design a Multi-Transverse-Mode Optical Processor (MTMOP) on 220 nm thick Silicon Photonics exploiting the first two quasi-transverse electric modes (TE0 and TE1). The objective is to measure the optical phase, required for programming the optical processor, without use of conventional optical phase detection techniques (*e.g.*, coherent detection). In the proposed design, we use a novel but simple building block that converts the optical phase to optical power. Mode TE0 carries the main optical signal while mode TE1 is for programming purposes. The MTMOP operation relies on the fact that the group velocity of TE0 and TE1 propagating through a *mode-sensitive* phase shifter are different. An unbalanced Mach-Zehnder interferometer (MZI) consists of a mode-sensitive and mode-insensitive phase shifters in the two arms. We set the bias of the phase shifters so that TE0 propagating in the two arms constructively interfere while this will not be the case for TE1. Hence, we detect the phase shift applied to TE0 by measuring the variation in the optical power of TE1. To the best of our knowledge, this design is the first attempt towards realizing a programmable optical processor with fully integrated programming unit exploiting multimode silicon photonics.

## I. INTRODUCTION

PROGRAMMABLE optical processors are promising structures for ultrafast and energy efficient optical computation. These processors can efficiently perform the vector-matrix multiplication portion of neural networks [1-4] from the inherent parallelism presents in optics in contrast with sequential operations in electronics [5]. Programmable optical processors can also pave the way for integrated microwave photonics (IMWP) [6], realize multiply-accumulate (MAC) operation in computing [7], and be used as in quantum computing [8]. With deep learning facing fast-growing computational demand limiting its progress [9-10], energy efficient computational accelerators fabricated in silicon photonic (SiPh) technology is a candidate to meet the computational demands of future machine learning and deep learning applications [11-14].

The programming techniques proposed for optical processors are mostly focused on *in-situ* training methods, where an optimization technique such as back propagation or gradient descent are used [15, 16]. These techniques require considerable amount of computation for programming an individual chip. Ideally, programmable optical processors should be fully reconfigurable by software after the fabrication like what is offered by the electronics field-programmable gate arrays (FPGAs) [17, 18]. One can perform *ex-situ* programming on an optical FPGA, *i.e.,* a specific weight matrix can be implemented on different similar chips. However, there are two main challenges in this approach. Firstly, programmable optical processors, unlike electronic FPGAs, are built on analogue building blocks more sensitive to the device parameters. Fabrication variations therefore translate into considerable computation error and accuracy in these processors [19-21]. Hardware error correction schemes are presented to tackle this issue [22]. Secondly, unlike *in-situ* training, *ex-situ* calibration and programming optical processors require sensing both the optical power and optical phase. Although sensing the optical power is easily feasible in photonic integrated circuits using on-chip photodetectors, sensing the optical phase requires more complex and elaborate hardware.

In this work, for the first time, we propose the ***Multi-Transverse-Mode Optical Processor (MTMOP)*** for large-scale optical computing applications. This is a novel architecture exploiting multiple quasi-transverse electric (TE) modes in an optical computation platform. The advantage of the MTMOP is its accurate, low-cost, and fast programming procedure capable of being integrated with SiPh. In a conventional programmable optical processor, one needs a coherent detector to measure the optical phase of phase shifters to program the processor [22]. Integrating a coherent detector in a SiPh chip increases the cost, area, power consumption, and complexity of the optical processor design [23]. In the MTMOP design presented in this work, we overcome this challenge by introducing a new but simple building block enabling optical phase measurement without the need for coherent detection. Using orthogonal TE optical modes, the MTMOP converts the optical phase into optical intensity that can then be easily measured on chip using

Manuscript submitted on 1 February 2022.
This work was supported in part by the Natural Sciences and Engineering Research Council of Canada (NSERC), [funding reference number DGDND-2021-03480]. K. R. Mojaver is holding Fonds de Recherche du Québec - Nature et technologies-B3X postdoctoral research scholarship, [funding reference number 302428].

K. Rahbardar Mojaver and O. Liboiron-Ladouceur are with the Department of Electrical and Computer Engineering, McGill University, Montreal, QC H3E 0E9, Canada. Corresponding authors: Kaveh Rahbardar Mojaver and Odile Liboiron-Ladouceur (e-mail: hassan.rahbardarmojaver@mcgill.ca; odile.liboiron-ladouceur@mcgill.ca).

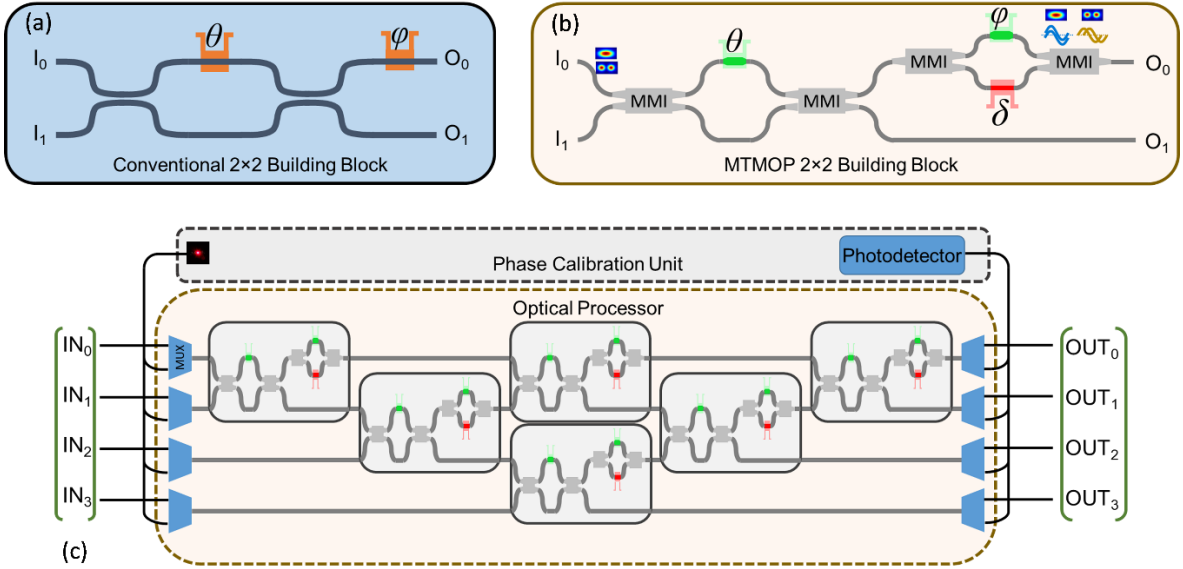

**Fig. 1.** (a) The 2 × 2 building block schematic of a conventional optical processor. (b) The 2 × 2 building block schematic of the MTMOP. We replace the external $\phi$ phase shifter by an MZI with a phase shift in each arm ($\phi$ and $\delta$). The phase shifter $\delta$ is mode sensitive with different thermo-optic coefficient for TE0 and TE1. The phase shifter $\phi$ is mode insensitive as the external phase shifter $\theta$. Optical mode TE0 constructively interferes through $\phi$ and $\delta$ while optical mode TE1 does not. Phase changes experienced by TE0 is measured through changes in the TE1 optical power. (c) A 4 × 4 MTMOP designed on a 4×4 Reck mesh.

optical photodetectors widely available in process design kit (PDK) of SiPh microfabrication foundries. The direct programmability is a big step towards on-chip programming of optical processors.

Section II provides the theory and background on the programmable optical processor. Section III present the working principle of the MTMOP. Section IV present the design and discussion. Conclusion is presented in section V.

## II. THEORY AND BACKGROUND

Figure 1 (a) shows the 2 × 2 building block for programmable optical processors. This block is a Mach-Zehnder interferometer (MZI) composed of two couplers and two phase shifters. The linear transformation matrix of the 2 × 2 building block for a fixed state of polarization, 50:50 splitting ratio of couplers and assuming lossless optical propagation is:

$$e^{j(\theta/2)} \begin{bmatrix} e^{j\varphi}\sin\left(\frac{\theta}{2}\right) & e^{j\varphi}\cos\left(\frac{\theta}{2}\right) \\ \cos\left(\frac{\theta}{2}\right) & -\sin\left(\frac{\theta}{2}\right) \end{bmatrix} \quad \text{eq. (1)}$$

where $\theta$ is the internal phase shift changing the output optical intensity, and $\phi$ is the external phase shift defining output optical phase [24].

In an ideal case considering no error, we can fabricate the 2 × 2 building block, characterize the phase shifters $\theta$ and $\phi$, and apply the same voltage-phase relation to every phase shifter on the chip. However, there are a few sources of errors in the phase shifter corrupting the relative phase/intensity requiring to calibrate and program each phase shifter individually. One can divide these errors into static and dynamic errors. Static errors are mainly due to the fabrication variations in the waveguide geometries or variations in the geometry of thermoelectric phase shifters. We can correct the static errors modifying the bias of the corresponding phase shifter during the first-time calibration process. Dynamic errors in phase shifters are more challenging to address. The main source of dynamic errors is thermal crosstalk between the phase shifters. One can minimize the thermal crosstalk by thermally isolating the phase shifters [25]. The second source of dynamic errors is the inaccuracy in the bias voltage/current applied to the phase shifter generated from digital-to-analogue converters (DACs) or even voltage drops through wire- bonding, pads, and on-chip electrical connections. Thermal drift is another source of dynamic error that can be compensated to a large extent by stabilizing the chip temperature using thermoelectric cooling (TEC) systems. A small variation of temperature leads to a small variation of optical phase change linearly dependent on temperature through temperature coefficient of a phase shifter. For example, a one Kelvin variation of temperature in a 100 µm long phase shifter with thermo-optic coefficient of $1.8 \times 10^{-4}$ K$^{-1}$ leads to 4.2 degree change of the phase shift at 1550 nm wavelength [26]. This small perturbation of phase can translate to a considerable shift of phase/intensity at the output due to the nonlinear nature of cascaded interferometers [24]. Therefore, small sources of errors usually neglected in SiPh integrated circuits designed for telecommunication applications cannot be neglected in optical processors applications. Besides phase shifters, beam splitters can also add to the hardware error of programmable optical processors. Bandyopadhyay et al. have demonstrated a programming method for readjusting the phase shifters' bias and correcting the splitting ratio error of beam splitters [22]. Optimization techniques such as *in-situ* back propagation training also compensate for this error [15, 16].

However, these techniques are not straightforward and are time consuming. Taking all these into account, there is a need for programming hardware capable of sensing both optical power and phase to set the bias of each phase shifter. One can calibrate and program the phase shifter $\theta$ by selecting a path including the corresponding MZI, set all the other MZI to either its minimum or maximum transmission ($\theta$ = 0 and $\pi$) and measuring the optical intensity at the output [22]. Sweeping $\theta$ generates a sinusoidal signal at the output optical power. To program/calibrate the external phase shifters (shown by $\phi$ in Fig. 1-a) we need a coherent detection to measure the optical phase since the phase shifter $\phi$ does not affect the amplitude of the optical signal and only affects its phase. The technique presented in this work converts the optical phase to optical power that can be measured using photodetectors.

### III. PRINCIPLE OF OPERATION

Figure 1 (b) presents the proposed $2 \times 2$ building block schematic of the MTMOP. It uses two orthogonal quasi-transverse electric (TE) optical modes: the fundamental quasi-transverse electric (TE0) mode for carrying the main optical signal and the first quasi-transverse electric (TE1) mode for performing phase calibration. The internal phase shifter $\theta$ is a mode-insensitive phase shifter applying the same phase shift to the TE0 and TE1 modes. In the MTMOP, we replace the external $\phi$ phase shifter by an MZI composed of I) a mode insensitive multimode interferometer (MMI) as a 50:50 beam splitter, II) two phase shifters ($\phi$ and $\delta$), III) a second mode insensitive MMI as the beam combiner. The phase shifter $\delta$ is mode sensitive with different thermo-optic coefficient ($dn_{eff}/dT$) for TE0 and TE1, where $n_{eff}$ is the effective refractive index and $T$ is the temperature. The phase shifter $\phi$ is mode insensitive.

Figure 1 (c) shows a $4 \times 4$ MTMOP in the Reck architecture [27] based on the aforementioned $2 \times 2$ building block. The idea presented in this paper can be extended to other programmable optical processor architectures such as Clement mesh [28] or diamond mesh [29]. The optical signal generated by the phase calibration unit along with the main optical signal ($IN_i$) are mode multiplexed and applied to the multi-mode input waveguide on TE1 and TE0, respectively. At the output we demultiplex the two modes. TE0 goes to the output and TE1 is detected by the phase calibration unit for the programming purpose.

The flowchart presented in Fig. 2 summarizes the procedure of programming a MTMOP. We start with the phase shifter $\theta$. Programming phase shifter $\theta$ is more straightforward since it defines the output optical power. We sweep the phase shifter $\theta$ voltage bias and measure the TE0 optical power at the top outputs ($O_0$ in Fig. 1 (b)). Considering the 50:50 splitting ratio of the splitter/combiner MMIs, the optical power is minimized and maximized at $\theta=0$ and $\theta=\pi$, respectively, for all values of $\phi$ and $\delta$.

For programming the external phase shifters, we start with setting a bias to $\phi$ as an initial point for the desired TE0 phase shift. We then sweep the phase shifter $\delta$ voltage bias until the TE0 signal power at $O_0$ is maximized meaning the TE0 signal passing through the phase shifters $\phi$ and $\delta$ constructively interferes. This would not be the case for TE1 owing to the mode sensitive nature of $\delta$. Knowing $dn_{eff}/dT$ for TE0 and TE1, we calculate the phase shift applied to TE0 by measuring the output amplitude of TE1. Knowing the phase shift applied to the TE0, we iterate the process until achieving the desired phase shift to TE0. Using this process, we can monitor the phase shift applied by the external phase shifters. Monitoring the phase shift is helpful in both calibration and programing phase of optical processors. We can calibrate the MTMOP completely on-chip using photodetectors and without need for coherent detection. Also, in the programming phase, the MTMOP enables monitoring the phase shift applied by the external phase shifter. This feedback signal can be used for closed loop programming and helps compensating the dynamic errors and achieving more accurate performance.

### IV. MTMOP DESIGN

The proposed MTMOP is designed for fabrication on a silicon-on-insulator (SOI) chip with a device thickness of 220 nm. The width of the waveguides for single mode propagation (TE0) and the multi-mode propagation (TE0 and TE1) are 0.43 μm and 0.96 μm, respectively. We use adiabatic directional coupler-based mode multiplexers (MUXs)/de-multiplexers (deMUXs) for mode conversion at the input and output. The design parameters for the MUXs/DeMUXs, MMIs, multimode S-bends are discussed in our previous works reported in [30, 31]. The phase shifters are thermo-optic phase shifters realized using high-resistance titanium-tungsten alloy (TiW). Contact with the heaters is made with a low-resistance titanium-tungsten/aluminum bi-layer (TiW/Al). As shown in Fig. 3, design of the phase shifters is done by simulating $dn_{eff}/dT$ for different TE modes as a function of the phase shifter's width using numerical tools. For the waveguide width larger than 4 μm, the difference between the values of the $dn_{eff}/dT$ for TE0 and TE1 is less than 1% and the phase shifter is thus *mode insensitive*. For smaller width of phase shifter, $dn_{eff}/dT$ varies for TE0 and TE1 resulting in a *mode sensitive* phase shifter. If we further decrease the phase shifter width below 0.96 μm, the propagation loss of TE1 drastically increases due to the overlap of its field distribution and the waveguide sidewalls. We select the width of 4 μm for mode insensitive phase shifters ($\theta$ and $\phi$) with $dn_{eff}/dT$=1.74. Further increase in the width of the mode insensitive phase shifter does not contribute considerably to the phase insensitivity, however, it decreases the power efficiency. We select a width of 0.96 μm for mode sensitive phase shifter ($\delta$) to maximize the phase sensitivity while maintaining low propagation loss for TE1 mode. For the phase shifter $\delta$, $dn_{eff}/dT$ is 1.8, and 1.96 for TE0 and TE1, respectively.

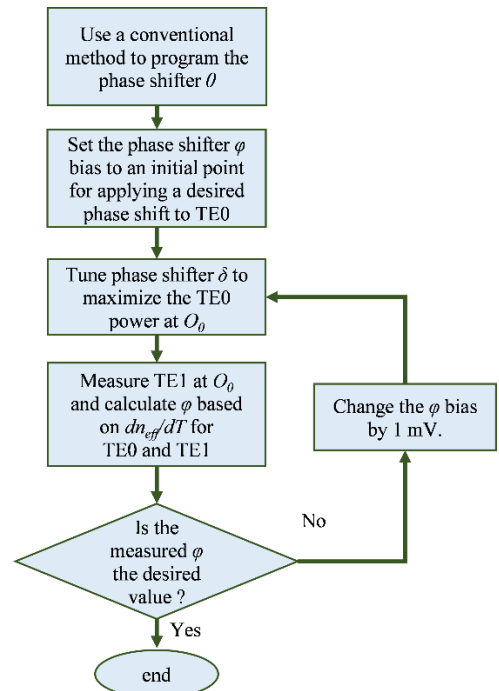

**Fig. 2.** Flowchart of the MTMOP programming procedure.

Figure 4 (a) plots the simulated TE0 phase shift passing through the phase shifters $\phi$ and $\delta$. As indicated in the flowchart of Fig. 2, we choose the bias of phase shifters $\phi$ and $\delta$ with the ratio to maintain constructive interference for TE0 at $O_0$. We scale the x-axis of all the plots in Fig. 4 to highlight this choice of bias for the two external phase shifters. Figure 4 (b) plots the simulated TE1 phase shift applied by the $\phi$ and $\delta$. The phase shifter $\phi$ applies the same phase shift to TE0 and TE1 due to its mode-insensitive characteristics. However, $\delta$ is mode-sensitive and gives different phase shift to TE1. Figure 4 (a) displays how the optical signal passing through $\phi$ and $\delta$ constructively interfere for TE0, while this is not the case for TE1 as shown in Fig. 4 (b). Figure 4 (c) plots the simulated TE0 output power (black dashed line), TE0 output phase (red dotted line), and TE1 output power (blue dash-dotted line) at $O_0$ versus the phase shifters $\phi$ and $\delta$ bias voltages. The TE0 and TE1 output power are normalized to their maximum value. While the optical output power of the TE0 mode remains constant, its output phase changes with the phase shift $\phi$. By monitoring the optical power change of TE1, the phase of TE0 can be inferred and we can therefore properly measure the phase shift applied to TE0 without the need for a coherent photodetection scheme. For example, if the TE1 power is -1.67 dB less from its maximum value, the corresponding phase shift of TE0 is 4π with a $\varphi$ bias of 3.1 V (Fig. 4c).

To achieve precise measurement of phase, a small change in $\phi$ should lead to a detectable change in the TE1 power. We define TE1 extinction ratio ($ER_{TE1}$) as the ratio of the TE1 optical power while phase $\phi$ changes by 2π. As shown in Fig. 4 (c), for $\phi$ between 0 and 2π, $ER_{TE1}$ is around 0.4 dB (Normalized TE1 is 0 dB for $\phi$=0 and -0.4 dB for $\phi$=2π). To increase $ER_{TE1}$ we can bias the phase shifter $\phi$ at larger values of voltage for example between 2.2 V and 3.1 V to get a phase shift of 2π to 4π. However, this approach leads to additional power consumption.

In this work, although $\delta$ is mode sensitive, $dn_{eff}/dT$ of this phase shifter is close in value for TE0 and TE1 (1.8 and 1.96, respectively, from Fig. 3), leading to an $ER_{TE1}$ that is relatively small in this design. We used a conventional narrow waveguide for the mode sensitive phase shifter with limited ability to demonstrate a proof of concept for the system level operation of MTMOP. Design of a more complex mode sensitive phase shifter with a larger difference in $dn_{eff}/dT$ of the optical modes would lead to a higher $ER_{TE1}$, and, thus, more dynamic range phase programmability. We define mode sensitivity ($\zeta$) for a phase shifter as the ratio of $dn_{eff}/dT$ for TE1 and TE0:

$$\zeta = \frac{dn_{eff}(TE1)/dT}{dn_{eff}(TE0)/dT} \quad \text{eq. (2)}$$

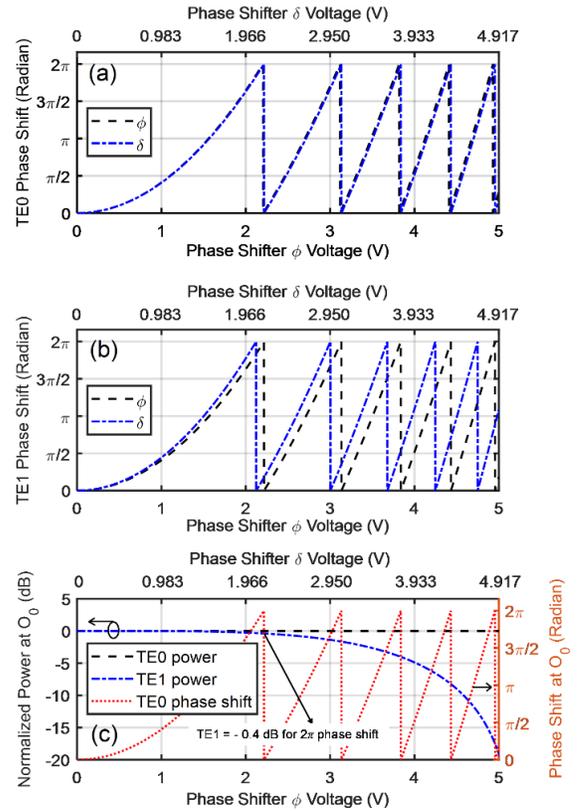

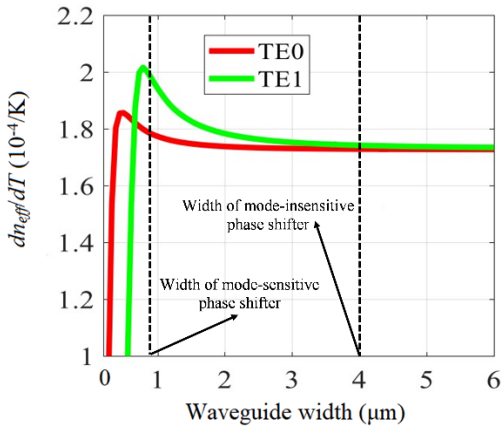

**Fig. 3.** Changes in the effective indices with temperature ($dn_{eff}/dT$) as a function of the phase shifter width for the first two TE modes.

**Fig. 4.** (a) TE0 phase shift applied by phase shifters $\phi$ and $\delta$, (b) TE1 phase shift applied by $\phi$ and $\delta$, (c) Output power of TE0 and TE1 at $O_0$, and phase shift applied to TE0 as functions of phase shifters voltage bias.

Figure 5 (a) shows the normalized TE1 optical power at $O_0$ versus the phase shift applied to TE0 for different values of $\zeta$. Following the procedure presented in the flowchart (Fig. 2), we maintain constructive interference at $O_0$ for TE0 by selecting an appropriate bias applied to $\delta$. As shown in Fig. 5 (a), increasing $\zeta$ from 1.09 (1.96/1.8) to 1.5 leads to a larger change in the detected TE1 power while sweeping $\phi$ from 0 to $2\pi$, thus larger $ER_{TE1}$. A mode sensitive phase shifter with $\zeta$ close to 1.5 can be realized using inverse design [32] and will highly contribute to an optimized performance of MTMOPs.

For $\zeta = 1.5$, $ER_{TE1}$ is maximized. In this case, for $2\pi$ phase shift applied to TE0 (*i.e.*, $2\pi$ accumulation for TE0 from both $\phi$ and $\delta$), the phase shift applied to TE1 is $2\pi$ in the $\phi$ arm (mode insensitive phase shifter) and $3\pi$ in the $\delta$ arm, leading to destructive interference for TE1. Therefore, the TE1 power at $O_0$ is minimum. For $\zeta > 1.5$, TE1 optical power would not be an injective (one-to-one) function of the phase shift applied to TE0 over $0 < \phi < 2\pi$ meaning that, for a single value of TE1 power one can read two values of phase shift. Thus, we must keep $\zeta \leq 1.5$ to estimate $\phi$ from TE1 optical power without requiring further analysis. Fig. 5 (b), shows the calculated $ER_{TE1}$ versus $\zeta$ of the phase shifter $\delta$. In this figure we show $\zeta > 1.5$ with a dash line to highlight the injective function part.

It is worth noting that the proposed MTMOP 2 × 2 building block includes an additional MZI compared to the conventional optical processors leading to a higher insertion loss. Also, the MTMOP uses multimode components (MMIs, waveguide bends, crossings, etc.) exhibiting more insertion loss compared to their single mode structure counterparts. Recently, there has been great improvement in developing SiPh multi-mode components to be used in mode-division-multiplexing (MDM) telecommunication systems [30, 31, 33]. Considering the developing trend in SiPh multi-mode components, MTMOP provides a viable solution to advance towards scalable self-programming optical processors.

In terms of the power consumption, the MTMOP includes an additional phase shifter per 2 × 2 building block compared to the conventional optical processor. This leads to an average increase of 50% power dissipation in the phase shifters. However, the MTMOP design provides a solution for on-chip monitoring of optical phase. This in turn reduces the complexity and elaboration of power-hungry hardware used for detecting the optical phase and simplify the optimization algorithms used for programming the optical processors.

## V. CONCLUSION

We propose the Multi-Transverse-Mode Optical Processor design for on-chip programming of optical processor. Exploiting the first two TE optical modes, the proposed design enables measurement of optical phase without use of coherent detection techniques. The group velocity of TE0 and TE1 propagating through a mode sensitive phase shifter is different, while both TE0 and TE1 modes travel with the same speed in a mode-insensitive phase shifter. In the unbalanced MZI with mode-sensitive and mode-insensitive phase shifters in the two arms, we can set the bias of the phase shifters so that TE0 mode constructively interfere while this will not be the case for TE1. We can then detect the phase shift applied to TE0 measuring the optical power change in TE1 eliminating the need for phase detection in the calibration of the optical processor.

## ACKNOWLEDGMENT

The authors would like to acknowledge the help of CMC Microsystems for subsidizing the applied nanotool (ANT) multi-project wafer fabrication run.

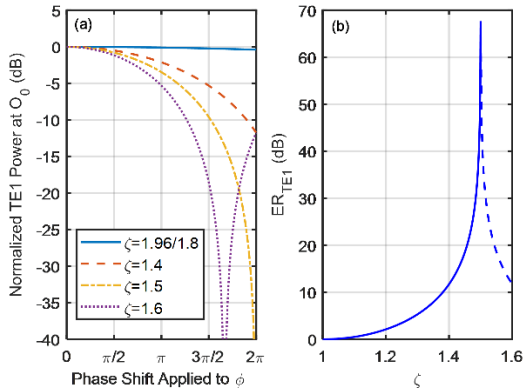

**Fig. 5.** (a) TE1 power at $O_0$ vs. phase shift applied to $\phi$ for different values of $\zeta$. (b) TE1 extinction ratio vs. $\zeta$.

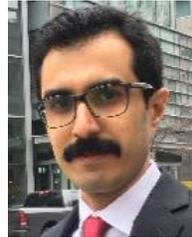

**Kaveh (Hassan) Rahbardar Mojaver** (M'17) received the B.S. and M.S. degrees in electrical engineering from the Amirkabir University of Technology (Tehran Polytechnic), Tehran, Iran, in 2009 and 2011, respectively, and the Ph.D. degree in electrical engineering from Concordia University, Montreal, QC, Canada, in 2018. He is currently a postdoctoral researcher with McGill University, Montreal. His research interests include development of photonics integrated circuits and systems for modern computing platforms, microfabrication, modeling and characterization of III-nitride HFETs. In 2019, he received the FRQNT-B3X postdoctoral research scholarship.

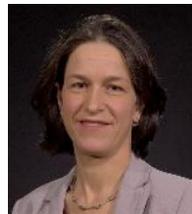

**Odile Liboiron-Ladouceur** (M'95, SM'14) received the B.Eng. degree in electrical engineering from McGill University, Montreal, QC, Canada, in 1999, and the M.S. and Ph.D. degrees in electrical engineering from Columbia University, New York, NY, USA, in 2003 and 2007, respectively. She is currently an Associate Professor and Canada Research Chair in Photonics Interconnect with the Department of Electrical and Computer Engineering, McGill University. She was an associate editor of the IEEE Photonics Letter (2009–2016) and was on the IEEE Photonics Society Board of Governance (2016–2018). She holds six issued U.S. patents and coauthored over 85 peer-reviewed journal papers and 125 papers in conference proceedings. She gave over 20 invited talks on the topic of photonic-electronic co-design, mode-division-multiplexing in silicon photonics, and optical interconnects. Her research interests include optical systems, photonic integrated circuits, optical neural networks, and mode-division multiplexing photonic interconnects. She is the 2018 recipient of McGill Principal's Prize for Outstanding Emerging Researcher.